\documentclass[10pt,conference]{IEEEtran}

\usepackage{booktabs} 
\usepackage[flushleft]{threeparttable}
\usepackage{xcolor}
\usepackage[utf8]{inputenc}
\usepackage{float}
\usepackage{amssymb}
\usepackage{amsmath}
\usepackage{ifthen}
\usepackage{caption}
\usepackage[bookmarks=false]{hyperref}
\usepackage{url,moreverb,xspace}
\usepackage[most]{tcolorbox}
\usepackage{enumitem}
\usepackage{array,graphicx}
\usepackage{soul}
\usepackage{balance}
\usepackage{pifont}
\usepackage[T1]{fontenc}
\usepackage{textcomp}
\usepackage{nicefrac}
\usepackage{mathtools}
\usepackage{tabularx}
\usepackage{xcolor,pifont}
\usepackage{tikz}
\usepackage{svg}
\usepackage{pdfpages}
\usepackage{array}
\usepackage{subcaption}
\usepackage{microtype}
\usepackage[cachedir=minted-cache]{minted}

\definecolor{mymauve}{rgb}{0.6,0,0.82}
\definecolor{lightgray}{gray}{0.9}
\definecolor{orange2}{rgb}{0.6,0,0.82}
\definecolor{mygreen}{rgb}{0.0,0.5,0.0}

\usepackage{listings}
\newcommand*\colourcheck[1]{%
  \expandafter\newcommand\csname #1check\endcsname{\textcolor{#1}{\ding{52}}}%
}
\newcommand*\colourcross[1]{%
  \expandafter\newcommand\csname #1cross\endcsname{\textcolor{#1}{\ding{56}}}%
}
\usepackage[vlined, boxruled, linesnumbered] {algorithm2e}
\SetKw{KwBy}{by}
\SetKw{KwBreak}{break}
\SetKw{KwReturn}{return}
\def\HiLi{\leavevmode\rlap{\hbox to \hsize{\color{gray!35}\leaders\hrule height .8\baselineskip depth .5ex\hfill}}}

\newlength\BARWIDTH
\newlength\BARHEIGHT
\SetKwComment{Comment}{$\triangleright$\ }{}
\SetCommentSty{itshape}
\newboolean{showcomments}
\setboolean{showcomments}{true}
\ifthenelse{\boolean{showcomments}}
{\newcommand{\nb}[2] {
  \fcolorbox{black}{gray!20}{\bfseries\sffamily\scriptsize#1:}
  {\sf\small$\blacktriangleright$\textit{#2}$\blacktriangleleft$}
}
}
{\newcommand{\nb}[2]{}
}

\newcommand{\head}[1]{\noindent\textbf{#1.}}

\newcounter{fcounter}
\newcommand{\curl}[1]{\footnote{\url{#1}}}

\makeatletter
\newcommand{\thickhline}{%
    \noalign {\ifnum 0=`}\fi \hrule height 1pt
    \futurelet \reserved@a \@xhline
}
\makeatother

\begin{document}

\title{Towards Automated Page Object Generation for Web Testing using Large Language Models}

\author{
\IEEEauthorblockN{Bet\"ul Karag\"oz\textsuperscript{1}, Filippo Ricca\textsuperscript{2}, Matteo Biagiola\textsuperscript{3}, Andrea Stocco\textsuperscript{1,4}}
\IEEEauthorblockA{\textsuperscript{1}\textit{Technical University of Munich}, Munich, Germany, {betul.karagoez, andrea.stocco}@tum.de}
\IEEEauthorblockA{\textsuperscript{2}\textit{Universit\`a degli Studi di Genova}, Genova, Italy, filippo.ricca@unige.it}
\IEEEauthorblockA{\textsuperscript{3}\textit{University of St.~Gallen and USI}, St.~Gallen / Lugano, Switzerland, matteo.biagiola@unisg.ch}
\IEEEauthorblockA{\textsuperscript{4}\textit{fortiss GmbH}, Munich, Germany, stocco@fortiss.org}
}

\IEEEoverridecommandlockouts

\maketitle

\IEEEpubidadjcol

\begin{abstract}
Page Objects (POs) are a widely adopted design pattern for improving the maintainability and scalability of automated end-to-end web tests. However, creating and maintaining POs is still largely a manual, labor-intensive activity, while automated solutions have seen limited practical adoption. In this context, the potential of Large Language Models (LLMs) for these tasks has remained largely unexplored. 
This paper presents an empirical study on the feasibility of using LLMs, specifically GPT-4o and DeepSeek Coder, to automatically generate POs for web testing. We evaluate the generated artifacts on an existing benchmark of five web applications for which manually written POs are available (the ground truth), focusing on accuracy (i.e., the proportion of ground truth elements correctly identified) and element recognition rate (i.e., the proportion of ground truth elements correctly identified or marked for modification). 
Our results show that LLMs can generate syntactically correct and functionally useful POs with accuracy values ranging from 32.6\% to 54.0\% and element recognition rate exceeding 70\% in most cases. Our study contributes the first systematic evaluation of LLMs strengths and open challenges for automated PO generation, and provides directions for further research on integrating LLMs into practical testing workflows.
\end{abstract}


\section{Introduction}\label{sec:introduction}

In modern software development, end-to-end (E2E) web testing is critical for ensuring the reliability of web applications. It validates complete user workflows by simulating real interactions through the browser, from logging in to completing complex tasks. Unlike unit or integration testing, which focus on isolated components, E2E web testing exercises the entire system in a production-like environment~\cite{2019-Ricca-Advances,2016-leotta-Advances}. 

Web automation frameworks such as Selenium~\cite{selenium}, Cypress~\cite{cypress}, and Playwright~\cite{playwright} have been developed to facilitate automated interactions between test scripts and web applications. 
These tools offer mechanisms to identify web elements and perform actions, such as clicking buttons and filling out forms, to enable automated test execution.
However, maintaining these tests remains a significant challenge: even minor modifications to the page layout, element identifiers, or navigation flow can cause tests to fail, resulting in fragile and costly maintenance efforts~\cite{2019-Ricca-Advances,ROBULAJournal,2018-Stocco-FSE}.

The Page Object (PO) design pattern is a common solution to improve test maintainability. 
By encapsulating web elements and their interactions in dedicated classes, the PO model decouples the test logic from the underlying GUI structure. 
This abstraction simplifies test management, reduces duplication, and allows for more robust scripts: if the GUI changes, only the corresponding PO needs to be updated. 
Moreover, POs can be reused across test suites, further improving maintainability~\cite{leotta2020family}. 
Despite these advantages, building and maintaining POs is labor-intensive, which hinders their widespread adoption.

To reduce this manual effort, prior research explored automated PO generation~\cite{po-lei-ma,2016-Stocco-SQJ,apogenclustering,StoccoLRT15,2020-Bajammal-TSE}.
While effective, these approaches do not use recent advances in Large Language Models (LLMs) in code generation and information processing, which are increasingly being integrated into software engineering workflows to automate coding tasks.
Unlike existing structural or clustering-based PO generators~\cite{po-lei-ma,2016-Stocco-SQJ,apogenclustering,StoccoLRT15}, which rely on static DOM parsing or handcrafted rules~\cite{2020-Yandrapally-ICSE}, this work explores the use of LLMs as generalizable PO synthesizers that can interpret markup, infer semantics, and produce executable Selenium code. 

This paper investigates specifically the effectiveness of GPT-4o and DeepSeek Coder to automate PO generation, selected for their complementary design goals and public accessibility. 
GPT-4o represents a general-purpose, instruction-tuned multimodal model optimized for reasoning and code synthesis across diverse programming contexts, while DeepSeek Coder is a specialized model explicitly trained on large-scale code corpora with strong emphasis on software engineering tasks. 
Together, they provide a balanced evaluation of both generalist and domain-specialized LLMs.

We study how LLMs interpret HTML structures, infer relationships between pages, and translate them into functional POs with syntactically correct and semantically meaningful methods. 
Using a benchmark of five web applications with manually written POs as ground truth, we evaluate the generated artifacts in terms of accuracy and element recognition rate. 

Our results show that both models can generate valid and reusable POs, with accuracy ranging from 32.6\% to 54.0\% and element recognition rate exceeding 70\% in most cases. 
While DeepSeek achieved slightly higher average accuracy and element recognition rate, differences with GPT-4o were not statistically significant. 
Both models consistently produced syntactically valid Java classes, but struggled with navigation methods and return types, where missing or incorrect definitions were the most frequent source of errors. 
Conversely, LLMs often generated extra elements and methods beyond the ground truth, some of which could potentially enrich test suites.

\noindent
This paper makes the following contributions:
\begin{itemize}
\item The first systematic empirical evidence on the effectiveness of LLMs for PO generation. Our empirical study evaluates the effectiveness of state-of-the-art LLMs (GPT-4o and DeepSeek Coder) in generating POs on a benchmark of five web applications. 
\item A replication package including the experimental library and dataset to support open research~\cite{replication-package}. 
\end{itemize}

\section{Background}\label{sec:background}



\begin{figure}[htb]
\centering
\subfloat[][\emph{Login page.}]
{\includegraphics[trim=0.1cm 0cm 0.1cm 0.1cm, clip=true, scale=0.27]{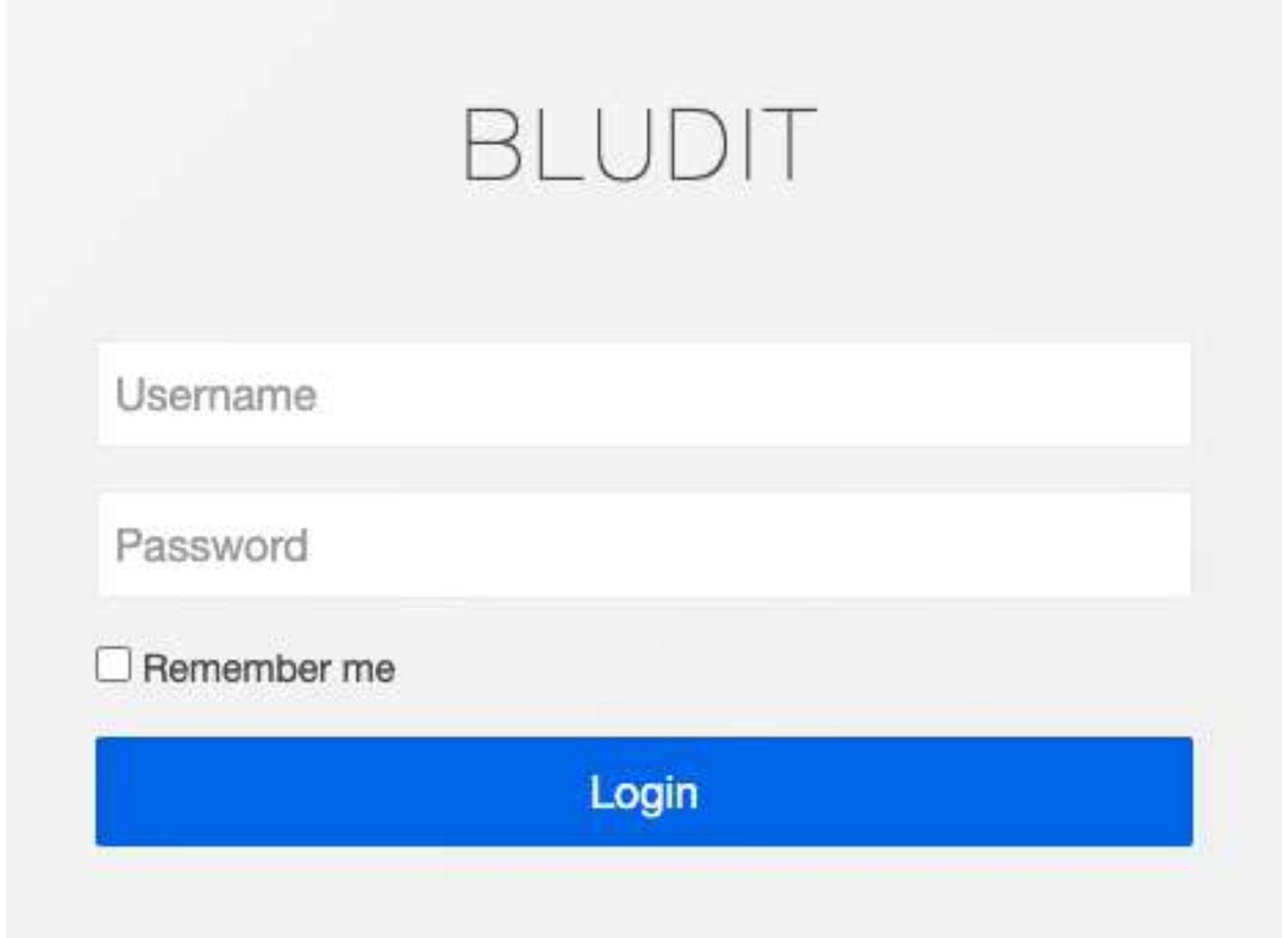}}
\quad
\subfloat[][\emph{Home page.}]
{\includegraphics[trim=0cm 10.5cm 0cm 7cm, clip=true, scale=0.4]{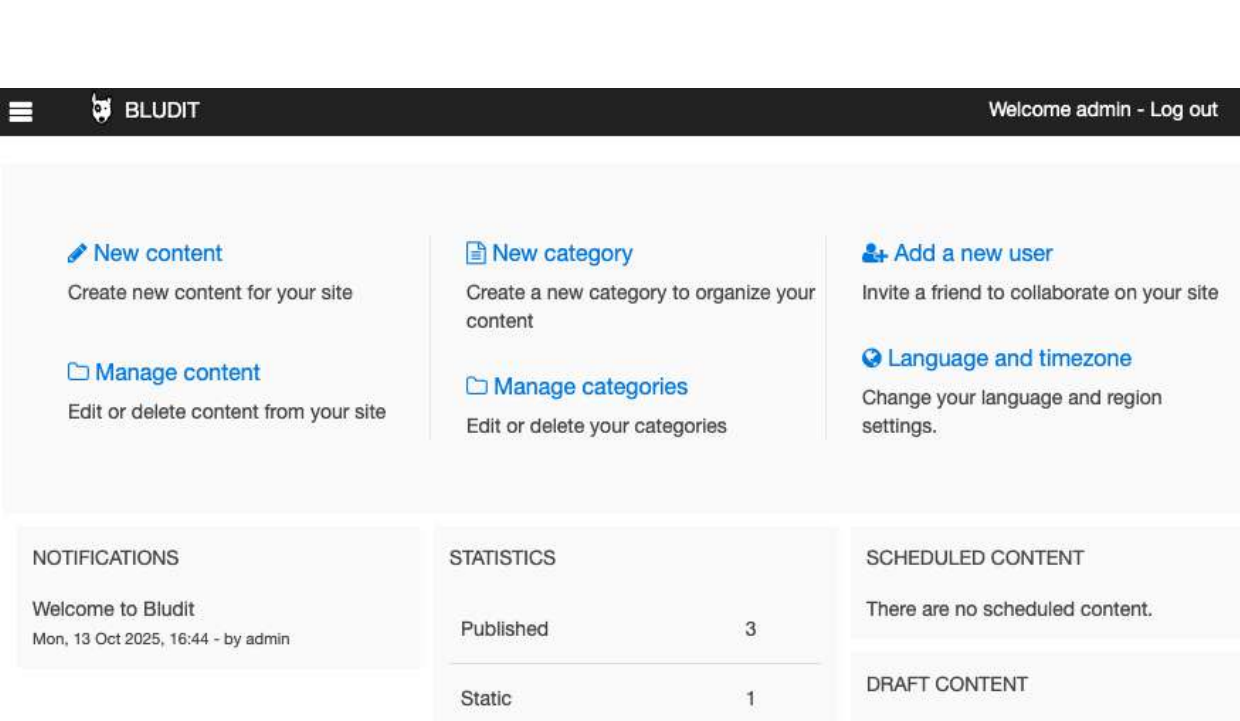}}
\caption{Bludit web application.}
\label{fig:ab-back}
\end{figure}

\autoref{fig:ab-back} shows our running example, i.e., an exceprt of the \texttt{Bludit} web application (v. 2.3.4).
We consider a scenario in which a user inserts \texttt{username} and \texttt{password} in the login form (\autoref{fig:ab-back}~\textbf{(a)}); if these credentials are correct, the \texttt{username} (in our example ``admin'') is displayed on  the top right corner of the homepage (\autoref{fig:ab-back}~\textbf{(b)}).

\subsection{Programmable Web Application Testing Tools} 
\label{sec:pttools}

With programmable web application testing tools, testers develop executable test cases using a high-level  programming language such as Java, Python, or Ruby. 
The tool APIs provide ways to interface with the browser; thus, test cases are composed of commands that either control the browser or provide input data, set the values of GUI components, and assert whether the program behaves correctly (e.g., by means of JUnit or TestNG assertions).

\autoref{lst:login-no-abs-back} displays a programmable version of a test case for the login functionality, created using Selenium WebDriver. 
In Line~3, an instance of WebDriver is created to control the Firefox browser. 
In Line~4, the executable test case instruments WebDriver to navigate to the URL of the target web application.  
Lines~5--10 fill in the username and password input boxes with the administrator credentials, and Lines~11--13 submit the form. 
Lines~14--16 verifies, by means of an assertion, the presence of an HTML element containing the username in the home page, and asserts that it equals the string ``admin''. 
Finally, Line~17 shuts down the WebDriver instance and closes the browser.

\definecolor{mymauve}{rgb}{0.6,0,0.82}
\definecolor{lightgray}{gray}{0.9}
\definecolor{orange2}{rgb}{0.6,0,0.82}
\definecolor{mygreen}{rgb}{0.0,0.5,0.0}


\begin{listing}[t]
    \begin{minted}[fontsize=\scriptsize,tabsize=4,linenos=true,numbersep=-5pt,autogobble,frame=lines,framesep=1em]{java}
        @Test
    	public void testLogin(){
    		WebDriver driver = new FirefoxDriver();
    		driver.get("localhost/Bludit/index.php");
    		driver.findElement(By
                .xpath("//*[@id=`LoginForm']/input[1]"))
                .sendKeys("admin");
    		driver.findElement(By
                .xpath("//*[@id=`LoginForm']/input[2]"))
                .sendKeys("secret");
    		driver.findElement(By
                .xpath("//input[@value=`Login']"))
                .click();
    		AssertThat(driver
                .findElement(By.id("loggedUser"))
                .getText(), is("admin"));
    		driver.quit();
    	}
    \end{minted}
    \caption{An example of a programmable automated test case.}
    \label{lst:login-no-abs-back}
\end{listing}

The implementation of advanced programmable test cases may require development skills 
comparable to those required for the development of production code. 
However, all of the benefits of modular programming can be brought to the creation of test suites, such as parametric and conditional execution, reuse of common functionalities across test cases  (e.g., applying design patterns), and robust mechanisms for referencing the HTML elements in a web page~\cite{ROBULAJournal}.

\subsection{Page Objects}\label{POPattern}

In web test automation, the ``Page Object'' design pattern~\cite{mf-po} has emerged as the leading pattern for enhancing test case maintenance, reducing code duplication and lowering the coupling between test cases and web applications~\cite{2016-Stocco-SQJ}. 
Page objects apply naturally to the programmable approach to web testing, because they are 
object-oriented classes that serve as interfaces to the web pages of the application under test. 
Test cases use the methods of the page object class whenever they need to interact with an HTML element of the GUI, which allows them to do anything that an end user can see and do on a page. 
A direct benefit of this is that test cases do not need to be modified if the application underneath undergoes structural changes~\cite{2016-leotta-Advances}, i.e., changes involving only the web page layout/structure. 
On the other hand, test cases still require maintenance in the presence of logical changes~\cite{2016-leotta-Advances}, i.e., those involving a web application's functionality.
There is empirical evidence of the benefits associated  with the adoption of the Page Object pattern in the maintenance of test suites for web applications, in both industry~\cite{Leotta-TAIC-2013} and academia~\cite{LeottaCRT13, leotta2020family}. 

\begin{listing}[t]
    \begin{minted}[fontsize=\scriptsize,tabsize=4,linenos=true,numbersep=-5pt,autogobble,frame=lines,framesep=1em]{java}
        public class IndexPage{
        	private WebDriver driver;
        	@FindBy(name="user")
        	private WebElement username;
        	@FindBy(name="pass")
        	private WebElement password;
        	@FindBy(id="submitLogin")
        	private WebElement login;
        	
        	public IndexPage(WebDriver driver){
        		this.driver = driver;
        		PageFactory.initElements(driver, this);
        	}
        	
        	public HomePage login(String usr, String pwd){
        		username.sendKeys(usr);
        		password.sendKeys(pwd);
        		login.click();
        		return new HomePage(driver);
        	}
    \end{minted}
    \caption{Java page object for the Login page in \autoref{fig:ab-back}~\textbf{(a)}.}
    \label{lst:po1}
\end{listing}

	
	

\begin{listing}[t]
    \begin{minted}[fontsize=\scriptsize,tabsize=4,linenos=true,numbersep=-5pt,autogobble,frame=lines,framesep=1em]{java}
        public class HomePage{
        	private WebDriver driver;
        	@FindBy(id="loggedUser")
        	private WebElement loggedUser;
        	
        	public HomePage(WebDriver driver){
        		this.driver = driver;
        		PageFactory.initElements(driver,  this);
        	}
        
        	public String getLoggedUser(){
        		return loggedUser.getText();
        	}
    \end{minted}
    \caption{Java  page object for the Home page of \autoref{fig:ab-back}~\textbf{(b)}.}
    \label{lst:po2}
\end{listing}

	


\autoref{lst:po1} and \autoref{lst:po2} present simplified page objects for \texttt{Bludit}. 
The classes are written in Java and implemented within the Selenium WebDriver framework. 
Each HTML element becomes a \texttt{WebElement} class instance\footnote{\url{https://seleniumhq.github.io/selenium/docs/api/java/org/openqa/selenium/WebElement.html}}, properly named and with a \texttt{@FindBy} annotation containing the locator. 
The page objects expose web page functionalities as methods; for simplicity, we limit the example to the methods of  interest to our sample login test case. 
\texttt{IndexPage} contains a method to perform the login, which wraps the calls to the 
various HTML elements in a single shareable unit (we assume the scenario in which credentials are correct, and that the result of the login operation is a page modeled with the \texttt{HomePage} page object).
The constructors of the page objects of \autoref{lst:po1} and \autoref{lst:po2} contain calls to a support library named \textit{Page Factory} (Line~12 and Line~8, respectively). 
The Page Factory instantiates the HTML elements of the page object and pre-populates them based on the annotations (e.g., Lines~3, 5 and 7 of \autoref{lst:po1}).
In the \texttt{HomePage} page object, the \texttt{getLoggedUser()} method (Lines~11--13 of \autoref{lst:po2}) retrieves the textual content associated with the HTML element having id equal to ``loggedUser''.

\begin{listing}[t]
    \begin{minted}[fontsize=\scriptsize,tabsize=4,linenos=true,numbersep=-5pt,autogobble,frame=lines,framesep=1em]{java}
        @Test
    	public void testLogin(){
    		IndexPage ip = new IndexPage(driver);
    		HomePage hp = ip.login("admin", "secret");
    		assertEquals(hp.getLoggedUser(), "admin");
    	}
    \end{minted}
    \caption{Page object-based programmable automated test case (Selenium WebDriver)}
    \label{lst:login-with-po}
\end{listing}


\autoref{lst:login-with-po} shows a version of the login test case originally presented in \autoref{lst:login-no-abs-back}, modified to use the page objects of \autoref{lst:po1} and \autoref{lst:po2}. 
The code has become more readable because the test case steps now correspond more directly 
to the steps of the test case specification (e.g., open the index page, log in to the 
application, and assert that the user is logged in).



\subsection{Benefits of Using Page Objects}

The main advantage of using the Page Object (PO) pattern is that changes to the user interface (UI) only require updates within the page object itself, not in the test scripts~\cite{leotta2023challenges}. 
All UI locators and page-specific actions are centralized in a dedicated page class, making maintenance easier when the UI evolves~\cite{leotta2020family}.

Another benefit is reusability. 
The same page objects can be used across multiple test cases. 
For example, a login operation (valid or invalid) can serve as a reusable action across various tests. 
This reduces code duplication and promotes a more modular, maintainable codebase.

The PO pattern also provides encapsulation and abstraction. 
Test scripts interact only with high-level methods, without needing to manage low-level UI details like locators~\cite{leotta2020family}. 
This makes it easier to expand the test suite as new pages or features are added to the application under test (AUT), requiring minimal changes to existing tests.
Additionally, POs enhance readability and clarity by exposing user actions through clearly named methods~\cite{neelapu2024enhancing}. 
This makes the tests easier to understand and maintain. 
Finally, debugging is simplified, as failures are isolated within the relevant page objects

\subsection{Challenges in Using Page Objects}

While POs offer clear benefits, such as decoupling, reduced code duplication, and easier test maintenance, Leotta et al.~\cite{leotta2023challenges} emphasize that creating POs for a web application is a nontrivial task. 
It requires significant development effort, particularly during the early stages of testing when POs must be built from scratch.
Although the advantages of adopting the PO pattern are well-documented, it remains uncertain whether these benefits always justify the upfront investment, especially for smaller-scale automation projects where POs may introduce unnecessary complexity.
Conversely, in large and dynamic applications, maintaining a growing set of page objects can become a burden, as highlighted by Stocco et al.~\cite{apogenclustering}.
\section{Study design}\label{sec:empirical-study}

This section describes the definition, the design, and the settings of the experiment we conducted in a structured way, following the guidelines by Wohlin et al.~\cite{wohlin00}.

\subsection{Goal and Research  Questions}\label{sec:rqs}
The goal of this study is to evaluate the effectiveness of Large Language Models in generating Page Objects (POs) for web testing, with a focus on their correctness and completeness, and to compare these results against manually written POs, which serve as the ground truth.
Our evaluation includes the following three research questions:

\noindent
\textbf{RQ\textsubscript{1} (Accuracy):} \textit{How effective are Large Language Models in generating Page Objects with high accuracy?}

\noindent
\textbf{RQ\textsubscript{2} (Element Recognition Rate):} \textit{To what extent are Large Language Models able to recognize ground-truth elements when generating Page Objects?}

\noindent
\textbf{RQ\textsubscript{3} (Issue categories):} \textit{In which specific categories of issues do Large Language Models struggle the most when generating Page Objects?}

The first research question investigates the overall accuracy of LLMs in generating POs, i.e., their ability to produce correct elements that match the ground truth. 
The second research question focuses on the element recognition rate, measuring to what extent LLMs are able to identify and reproduce existing elements in the target application. 
The third research question analyzes the types of issues that arise in LLM-generated POs, with the goal of understanding where these models most frequently fail. 

\subsection{Objects of the Study}

%
In our study, we evaluate LLM-generated POs by comparing them with manually created POs. For this reason, the Applications Under Test (AUTs) must already have manually written POs available. To ensure consistency and reliability, we used the \texttt{BEWT} benchmark of web applications~\cite{olianas2025bewt}, which provides developer-written POs, Docker containers to run the applications, and associated test cases. From this benchmark, we selected five applications from different categories, allowing us to assess LLMs performance across diverse scenarios, detailed in \autoref{tab:app-stats} and described next.

\texttt{Bludit} (v. 3.13.1) is lightweight content management system (CMS) that allows users to create and manage websites. It does not require a database, but instead it relies on flat files~\cite{bludit}. Its simplicity and fast performance makes it a useful test application for evaluating LLMs in content-oriented applications. 
\texttt{ExpressCart} (v. 1.19) is an open-source e-commerce application built with Node.js. It includes main features like listing products, shopping cart functionality, and checkout processes~\cite{expresscart}. 
\texttt{Kanboard} (v. 1.2.15) is a project management tool which uses the Kanban methodology to help users visualize tasks, manage workflows, and track project progress~\cite{kanboard}. 
\texttt{MediaWiki} (v. 1.40.0) is a collaborative wiki platform best known for powering Wikipedia~\cite{wikipedia}. It supports structured content, revision history, and user permissions~\cite{mediawiki}. 
\texttt{PrestaShop} (v. 1.7.8.5) is an open source e-commerce platform that provides a set of functionalities for managing online stores. These functions include product management, order processing and customer interactions~\cite{prestashop}.

\begin{table}[t]
\centering
\caption{Statistics of the Benchmark Applications.}
\resizebox{\columnwidth}{!}{
\begin{tabular}{lrrrr}
\toprule
\textbf{Application} & \textbf{Source LoC} & \textbf{Test LoC} & \textbf{\#Test Cases/POs} & \textbf{Version} \\
\midrule
\texttt{Bludit}      & $\sim$12{,}000   & $\sim$500    & $\sim$50   & 3.13.1 \\
\texttt{ExpressCart} & $\sim$8{,}500    & $\sim$300    & $\sim$40   & 1.19 \\
\texttt{Kanboard}    & $\sim$25{,}000   & $\sim$3{,}000 & $\sim$120  & 1.2.15 \\
\texttt{MediaWiki}   & $\sim$150{,}000  & $\sim$10{,}000 & $\sim$300  & 1.40.0 \\
\texttt{PrestaShop}  & $\sim$220{,}000  & $\sim$15{,}000 & $\sim$400  & 1.7.8.5 \\
\bottomrule
\end{tabular}
}
\label{tab:app-stats}
\end{table}

\subsection{Benchmark Setup}

The \texttt{BEWT} benchmark includes POs, Docker containers for running web applications, and test scripts. 
However, it does not provide the HTML content, which is the primary input required for prompting LLMs. To gather the necessary HTML for the screens, we ran the applications locally using the provided Docker containers, navigated to the relevant pages, and saved the corresponding HTML files. These raw HTML files, however, often contain excessive or redundant data, such as scripts and embedded resources, that could easily overload the context for LLMs~\cite{huq2023whats, liu2024lost}.
To address this issue, we implemented a pre-processing step to truncate and clean the HTML content before feeding it into the model. This approach was informed by the findings of Huq et al.~\cite{huq2023whats}, which showed that feeding raw HTML files to LLMs often leads to hallucinations and failure to follow instructions. Specifically, the study found that applying an effective truncation strategy could improve model performance by up to 11\%.

\subsection{Selection of Large Language Models}

We selected GPT-4o~\cite{openai2024gpt4o} and DeepSeek Coder~\cite{deepseek2024llm} (DeepSeek, hereafter) for our experiment on automated PO generation due to their state-of-the-art performance and strong specialization in code-related tasks. Initially, our goal was to rely on smaller, open-source models for accessibility and lower resource demands, but preliminary tests showed they lacked sufficient accuracy. Larger open-source models performed better but exceeded our computational capacity. Therefore, we focused on high-performing models available via external APIs, which allowed us to evaluate advanced capabilities without the burden of running large models locally.



\subsection{Large Language Model Setup}

\subsubsection{Hyperparameters}

LLMs contain various hyperparameters that change the way they behave, which ends up affecting the output quality. 
Among these, there is the \textit{temperature parameter}, which controls the creativity of the output. Higher values (closer to 1) generate more diverse responses, while lower values (closer to 0) produce more consistent and deterministic results. Since the task requires consistency rather than creativity, we set the temperature to 0.1 in all experiments, mitigating the impact of randomness in the results~\cite{ouyang2025empirical}.

Another important hyperparameter is \textit{max tokens}, which defines the maximum number of tokens that can be generated in a single interaction by the LLM. Since our task requires a complex output, we chose to use the default value (i.e., 16,384 for GPT-4o and 2,048 for DeepSeek Coder).

\subsubsection{Prompt Strategies}

Previous studies show that prompts are crucial in guiding LLM performance~\cite{brown2020language}. 
Clear, well-structured prompts define the task and lead to good results, while vague prompts cause confusion and off-target outputs. 


%

In our approach, we assume that no existing test suites or pre-defined POs are available for the selected AUT. This leads us to adopt a zero-shot prompting approach, which does not rely on prior examples or predefined structures. Instead, it allows the model to generate POs and test cases based solely on the provided requirements or task description.
This approach enables us to evaluate an LLM's ability to create correct POs without any prior knowledge of the AUT. It also serves as a means to assess the model's capabilities in real-world scenarios, where no setup or input from existing resources is available.
Being the first study of this kind, we adopted a zero-shot setup to establish a reproducible lower bound of model capability, isolating intrinsic LLM reasoning from external aids such as few-shot examples or retrieval. 

\autoref{lst:page-object-prompt} shows the prompt structure we used to specify the rules for generating POs together with the corresponding HTML content for the page. The LLM is expected to generate a PO in Java, adhering to a specific structure outlined in the prompt. This includes using appropriate naming conventions, setting up the constructor, and creating \texttt{WebElements} for each input field, button, and link on the page. The generated PO must also import necessary libraries from Selenium and JUnit, which we use in this work. Additionally, it should include methods for navigation (e.g., clicking buttons or links) and interactions (e.g., entering text).

During our experiments, we noticed that some requirements were not implemented, even though they were clearly stated in the prompt. Upon further analysis, we found that using more assertive language—specifically imperative statements like ``must'', increased the likelihood of those requirements being correctly implemented. Consequently, we refined the prompt by using ``must'' statements to clearly highlight critical instructions, improving model compliance.



\begin{lstlisting}[caption={Page Object Generation Prompt.}, label={lst:page-object-prompt}, numbers=none]
You are generating a Selenium Page Object Model (POM) class in Java for a web application.
    
### File Naming Rules:
- The Java class must be named **exactly**: {class_name}
    - This name must be used:
        - As the Java **filename** (`{class_name}.java`) **(must match exactly)**
        - As the Java **class name** (`public class {class_name}`) **(must match exactly)**
        - Do not infer or modify the class name based on page content
    
### Generation Instructions:
    1. You must define all key elements (inputs, buttons, links, text fields, etc.) as `WebElement` fields using `@FindBy`annotations.
    2. At the top of the class, you must include this field:
        `public WebDriver driver;`
    3. You must use **only generic, structural field names**, based on structure 
    or position:
        - Examples: `firstButton`, `mainInputField`, `headerLabel`, `secondaryLink`
        - You must **not** use domain-specific terms like "product", "cart", "price", etc.
    4. If multiple similar elements are present (like multiple buttons or links), you should use **ordinal names**: `firstButton`, secondButton`, etc. 
    (optional unless needed for clarity).
    5. You must use **camelCase** for all field and method names.
    6. Create interaction methods that:
        - Perform actions like clicking, typing, or retrieving text.
        - If clicking an element clearly leads to another page 
        (e.g., a link, a save button after form submission), **the method must return the appropriate POM class** (assume it already exists).
            - Example: after clicking a save button, return new ProjectSummaryPage(driver);`
        - If the action does not cause page navigation (typing into inputs, basic clicks), the method must return `void` or the appropriate text (`String`) if retrieving values.
    7. You must add a constructor with:
        `PageFactory.initElements(driver, this);`
    8. You must not generate any other classes unless explicitly instructed.
    9. You must only use information available in the provided HTML, do not assume additional logic or elements
    
### Error Message Handling Rules:
    - If the HTML contains elements indicating errors, validation messages, 
    or warnings (e.g., classes or IDs like `error`, `form-errors`, 
    `validation-error`, `alert-danger`):
        - You must create a `WebElement` field for each error message.
        - You must create a method named `getErrorMessage()` that returns the 
        visible error text (`element.getText()`).
        - If multiple error messages exist, you must create methods like `getFirstErrorMessage()`, `getSecondErrorMessage()`, etc., or alternatively return a list of error texts if appropriate.

### HTML Content:
    {html_content}
        """
\end{lstlisting}

\begin{table*}[t]
\centering
\caption{Detailed classification of Page Object review using GPT-4o and DeepSeek, for all apps. 
}
\resizebox{\textwidth}{!}{%
\begin{tabular}{
  l
  @{\hspace{20pt}}cccc
  @{\hspace{20pt}}cccc
  @{\hspace{20pt}}cccc
  @{\hspace{20pt}}cccc
}
\toprule
\textbf{App} 
& \multicolumn{4}{c}{\textbf{Correct}} 
& \multicolumn{4}{c}{\textbf{To Modify}} 
& \multicolumn{4}{c}{\textbf{Missing}} 
& \multicolumn{4}{c}{\textbf{Extra}} \\
\cmidrule(lr){2-5} \cmidrule(lr){6-9} \cmidrule(lr){10-13} \cmidrule(lr){14-17}
& Elem & Get & Act & Nav 
& Elem & Get & Act & Nav 
& Elem & Get & Act & Nav 
& Elem & Get & Act & Nav \\
\midrule
Bludit - GPT-4o       & 46 & 3 & 0 & 1   & 0 & 0 & 26 & 24   & 12 & 15 & 5 & 7     & 106 & 23 & 47 & 22 \\
Bludit - Deepseek     & 48 & 9 & 0 & 3   & 0 & 1 & 23 & 19   & 10 & 8  & 7 & 11    & 110 & 51 & 46 & 26 \\
Kanboard - GPT-4o     & 46 & 2 & 1 & 6   & 4 & 3 & 20 & 15   & 19 & 19 & 3 & 5     & 178 & 16 & 70 & 91 \\
Kanboard - Deepseek   & 48 & 10 & 2 & 4  & 7 & 3 & 17 & 20   & 14 & 11 & 5 & 2     & 253 & 44 & 82 & 95 \\
MediaWiki - GPT-4o    & 30 & 2 & 2 & 0   & 3 & 0 & 22 & 11   & 8  & 7  & 6 & 7     & 88  & 6  & 44 & 30 \\
MediaWiki - Deepseek  & 28 & 3 & 0 & 1   & 2 & 0 & 23 & 4    & 11 & 6  & 7 & 13    & 142 & 15 & 86 & 14 \\
PrestaShop - GPT-4o   & 73 & 5 & 0 & 3   & 4 & 0 & 4  & 25   & 30 & 14 & 9 & 11    & 212 & 6  & 145 & 74 \\
PrestaShop - Deepseek & 69 & 8 & 1 & 1   & 4 & 0 & 3  & 23   & 34 & 11 & 9 & 15    & 235 & 26 & 178 & 36 \\
ExpressCart - GPT-4o  & 20 & 1 & 2 & 2   & 0 & 3 & 6  & 11   & 2  & 1  & 0 & 2     & 69  & 9  & 41 & 13 \\
ExpressCart - DeepSeek& 21 & 3 & 2 & 1   & 0 & 0 & 8  & 12   & 1  & 0  & 0 & 2     & 77  & 19 & 39 & 9 \\
\midrule
\textbf{Total} & \textbf{429} & \textbf{46} & \textbf{10} & \textbf{22} & \textbf{24} & \textbf{10} & \textbf{152} & \textbf{164} & \textbf{141} & \textbf{92} & \textbf{51} & \textbf{75} & \textbf{1470} & \textbf{215} & \textbf{778} & \textbf{410} \\
\bottomrule
\end{tabular}%
}
\caption*{\footnotesize Abbreviations: Elem~$=$ Web Elements; Get~$=$ Getter Methods; Act~$=$ Action Methods;  Nav~$=$ Navigation Methods.}
\label{tab:po_detailed_review_total}
\end{table*}

\subsection{Evaluation Setup}

In the evaluation phase of our study, we manually compared the ground-truth POs provided in the benchmark with the outputs generated by both GPT4-o and DeepSeek across five different applications. 
For each generated POs, we manually compared the \texttt{WebElement} objects (e.g., username and password \texttt{WebElement}) and corresponding methods (e.g., login method) against the ground truth. In order to ensure consistency and clarity in our analysis, we categorized each identified element and method into four main groups, namely correct, to modify, missing, and extra, following prior work~\cite{StoccoLRT15}.

\head{Correct} Web elements are classified as \textit{correct} if they use the appropriate element type and if their assigned name is meaningful and matches, or closely aligns with, the ground truth. For methods, we require the method name to be accurate, the implementation to align with the intended behavior, and the return type to match the one defined in the ground-truth version, to be classified in this category.
    
\head{To modify} If an element is present but does not fully follow the naming conventions or structural guidelines given in the prompt, it is categorized as an element which requires modification by a developer. These elements are partially correct, and can still be used in the POs with manual adjustments. An important part of this classification was identifying specific shortcomings, which helped us detect recurring patterns in how the LLM misinterpreted or failed to capture the complete structure of certain elements. If the element marked for modification is not a method, we check its naming and type to determine in which part adjustments are needed. However, if the element is a method, we perform a more detailed analysis based on the requested fix. 

\head{Missing} Elements are classified as \textit{missing} if they are completely absent from the generated output. This applies to both web elements and methods that exist in the ground truth but are not generated by the LLM;

\head{Extra} Elements are categorized as \textit{extra} if they appear in the LLM-generated POs but have no corresponding match in the ground-truth POs. Since manually created POs may not always be complete or fully representative of the page's functionality, extra elements generated by LLMs reflect the model's ability to generate more comprehensive POs, potentially capturing additional functionality that may have been missed in the manually created ones.

The task was performed by the first author and validated through discussions with the other authors to resolve disagreements in ambiguous cases. No method was left unassigned.

\subsection{Metrics and Analysis for Evaluation}
To evaluate the performance of LLMs in automating PO creation, we used the following metrics.
For RQ\textsubscript{1}, we computed the accuracy metric is used to measure the percentage of methods that are classified as ``Correct''. In particular, accuracy is defined as follows: 

{
    \[
        \text{Accuracy} = \frac{\text{No. of Correct Methods}}{\text{Total Number of Methods}} \times 100
    \]
}
The corresponding metric for web elements is defined analogously.
For RQ\textsubscript{2}, this metric measures the proportion of ground truth elements that are successfully classified as ``Correct'' or ``To modify''. This will give us an idea of how much of the ground truth LLMs are able to accurately process and identify for potential change:

    {
    \[
        \text{Elem Rec. Rate} = \frac{\text{Total Correct + Requires Mod. Elems}}{\text{Total Elements}} \times 100
    \]
    }

The corresponding metric for methods is defined analogously.
For RQ\textsubscript{3}, we broke down the results of the ``To Modify'' category into specific subcategories. These subcategories represent specific types of issues where the LLM output does not match the ground-truth POs. For each subcategory me measure:

    {
    \[
        \text{Subcategory \%} = \frac{\text{No. of Elements in Subcategory}}{\text{Total Number of Elements to Modify}} \times 100
    \]}

The task was performed by the first author and validated through discussions with the other authors with prior experience in web testing to resolve disagreements in ambiguous cases, following the qualitative analysis protocol of Wohlin et al.~\cite{wohlin00}. No case was left unassigned.
The corresponding metrics for web elements/methods are defined analogously to those used for methods/web elements. This triangulation is aimed at mitigating subjectivity in class assignments.

\section{Results}


\subsection{RQ\textsubscript{1} (Accuracy)}

\autoref{tab:po_detailed_review_total} provides an overview of how accurately each LLM generated POs compared to the ground truth. We focus here on elements classified as \textit{Correct}, which reflect full alignment in type, naming, and behavioural semantics.

Across all applications, the majority of correctly generated items were WebElements (429 in total), indicating that LLMs are fairly reliable in detecting static page components. In contrast, methods, particularly Action and Navigation methods, were identified less accurately, confirming that behavioural elements remain challenging due to their dependency on application flow~\cite{2020-Biagiola-ICST}. Between the two models, DeepSeek showed marginally better performance than GPT-4o, especially in correctly generating Getter methods. However, neither model exceeded an overall accuracy threshold sufficient for full automation without developer intervention.

\begin{figure}[t]
  \centering
  \includegraphics[trim=0.1cm 8cm 0.1cm 8cm, clip=true, scale=0.42]{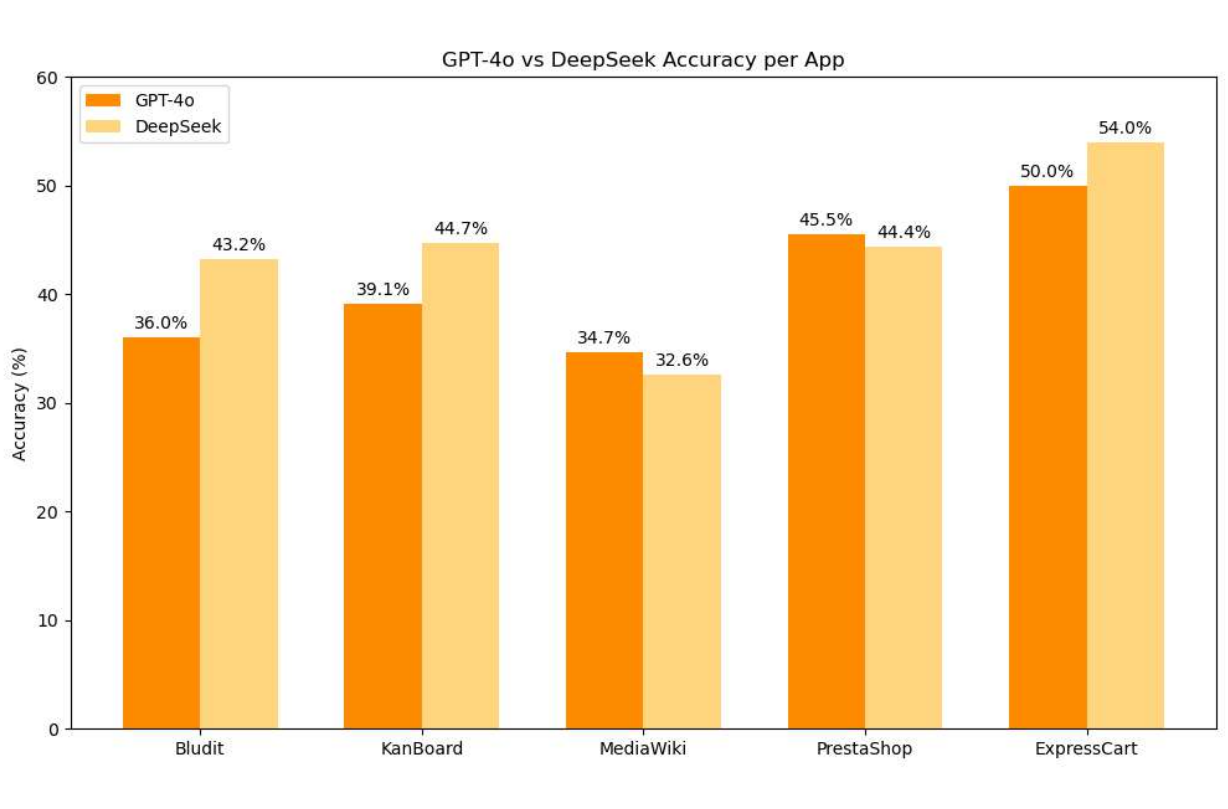}
  \caption{Accuracy of GPT-4o and DeepSeek.} 
  \label{fig:accuracy}
\end{figure}

\autoref{fig:accuracy} presents the overall accuracy comparison across all applications for both LLMs. 
In three of the five applications, Bludit, KanBoard and ExpressCart, DeepSeek outperformed GPT-4o and achieved higher accuracy scores. 
On the other hand, GPT-4o performed slightly better than DeepSeek in MediaWiki and PrestaShop applications. 
On our data, the results suggest that DeepSeek performs slightly better in terms of accuracy for PO creation. The lowest results were achieved in the MediaWiki application with accuracy values of 34.7\% and 32.6\% for GPT-4o and DeepSeek, respectively. 
On the other hand, the highest accuracy values were obtained in the ExpressCart application with 50\% in GPT-4o and 54\% in DeepSeek.

\begin{tcolorbox}[boxrule=0pt,frame hidden,sharp corners,enhanced,borderline north={1pt}{0pt}{black},borderline south={1pt}{0pt}{black},boxsep=2pt,left=2pt,right=2pt,top=2.5pt,bottom=2pt]
\textbf{RQ\textsubscript{1} (Accuracy):} LLMs can generate syntactically valid and partially functional POs, achieving accuracy between 32.7\% and 54.0\%. DeepSeek slightly outperforms GPT-4o in method-level correctness.
\end{tcolorbox}

\subsection{RQ\textsubscript{2} (Element Recognition Rate)}

For RQ\textsubscript{2}, we analyse the extent to which LLMs correctly detect relevant elements, regardless of whether they require refinement. Here, we consider both \textit{Correct} and \textit{To Modify} instances as successful recognitions.

The element recognition rate (\autoref{fig:coverage}) ranging from 61.2\% to 94\%, indicates that, even when LLMs do not produce fully correct implementations, they are generally capable of identifying the presence and role of page components. 
DeepSeek again demonstrates slightly higher recognition capabilities, particularly in complex applications such as Kanboard and MediaWiki, where structural depth poses additional challenges. However, both models frequently introduce \textit{Extra} elements, suggesting a tendency to overgenerate or infer functionality beyond the ground truth.

\begin{tcolorbox}[boxrule=0pt,frame hidden,sharp corners,enhanced,borderline north={1pt}{0pt}{black},borderline south={1pt}{0pt}{black},boxsep=2pt,left=2pt,right=2pt,top=2.5pt,bottom=2pt]
\textbf{RQ\textsubscript{2} (Element Recognition Rate):} LLMs reliably detect most relevant page elements, with element recognition rates between 61.2\% and 94\%. DeepSeek shows a modest advantage in recognising method-level components.
\end{tcolorbox}

\begin{table}[t]
\centering
\scriptsize
\renewcommand{\arraystretch}{1}
\caption{Breakdown of ``To Modify'' Subcategories.}
\begin{tabular}{@{}lccccc@{}}
\toprule
\textbf{LLM} & \textbf{Miss. Ret.} & \textbf{Wrong Ret.} & \textbf{SamePg Ret.} & \textbf{Wrong Nm.} & \textbf{Miss. Im.} \\
\midrule
GPT-4o    & 54 & 14 & 74 & 33 & 2 \\
DeepSeek  & 54 & 12 & 67 & 28 & 8 \\
\bottomrule
\end{tabular}
\caption*{\footnotesize Abbreviations: Miss. Ret.~$=$ Missing Return Type; Wrong Ret.~$=$ Wrong Return Type; SamePg Ret.~$=$ Same Page Return Type; Wrong Nm.~$=$ Wrong Name; Miss. Im.~$=$ Missing Implementation.}
\label{tab:breakdown_to+modify}
\end{table}

\subsection{RQ\textsubscript{3} (Issue categories)}

\autoref{tab:breakdown_to+modify} shows the total number of instances across all applications for issues that occurred in the ``To modify'' category. Our analysis revealed the following five categories:

    \begin{itemize}
        \item \textbf{Missing return type}: The method lacks an explicitly defined return type;
        \item \textbf{Wrong return type}: The method includes a return type, but it does not match the expected one;
        \item \textbf{Missing return type (returns same page)}: The method returns void, but according to the ground truth, it should have returned the same PO to support chaining in test cases~\cite{biagiola2017search};
        \item \textbf{Wrong naming}: The method name does not match the naming rules or does not reflect its intended functionality;
        \item \textbf{Wrong implementation}: The method logic is incorrect or differs from the expected behavior defined in the ground truth;
        \item \textbf{Missing implementation}: The method signature is defined but the actual implementation is left empty, which makes the method non-functional.
    \end{itemize}

It can be seen that the largest proportion of these issues were related to missing same page return type in both LLMs, with 74 out of 181 instances in GPT-4o, and 67 out of 169 instances in DeepSeek, with corresponding percentages of 40.9\% and 39.6\% respectively. Both LLMs had the same number of completely missing return type issues with 54 instances each. GPT-4o showed slightly more incorrect return types (14) and a higher number of naming issues (33 vs. 28) compared to DeepSeek (12), indicating a slightly higher tendency towards semantic inconsistencies in method identification. Finally, DeepSeek showed a higher number of missing implementations (8 compared to GPT-4o's 2), suggesting that it occasionally omits parts of the expected method logic. 

\begin{figure}[t]
  \centering
  \includegraphics[trim=0.1cm 8cm 0.1cm 8cm, clip=true, scale=0.42]{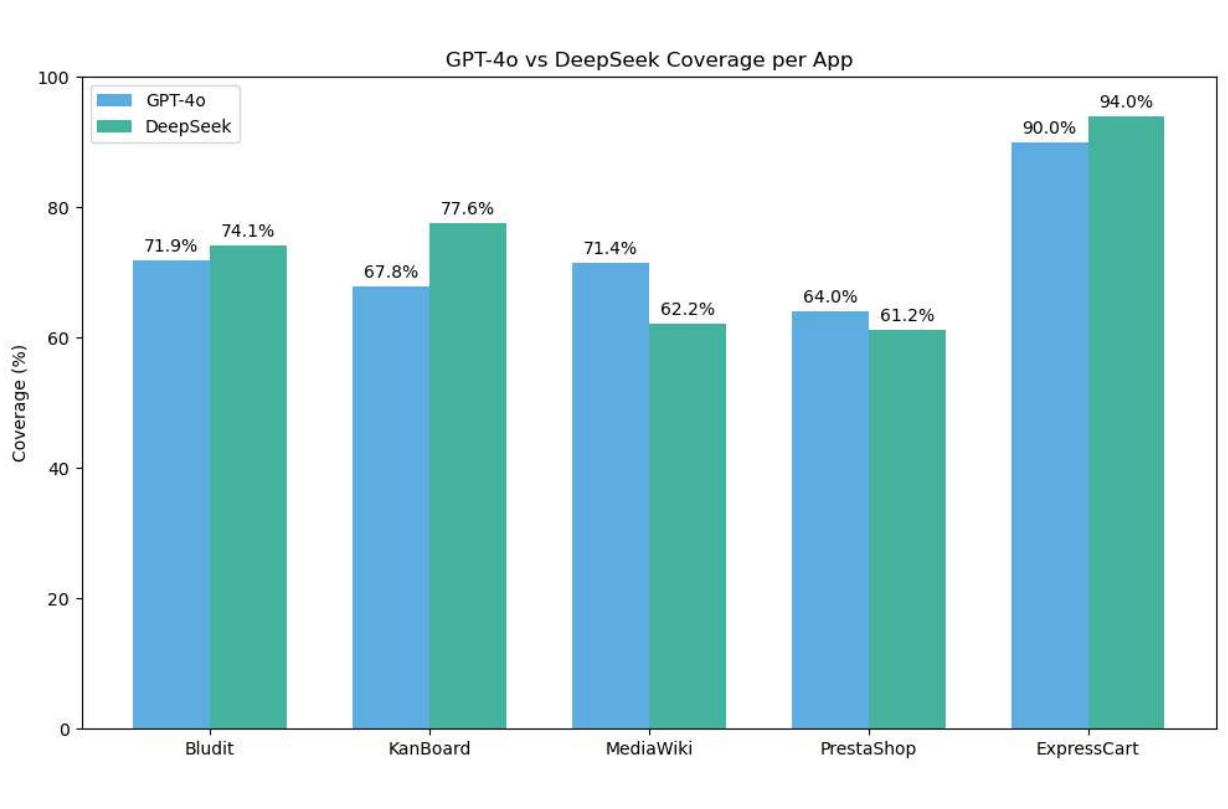}
  \caption{Element Recognition Rate of GPT-4o and DeepSeek.} 
  \label{fig:coverage}
\end{figure}

\autoref{fig:coverage} shows the overall element recognition rate comparison across all applications for both LLMs. Similar to the accuracy results, DeepSeek outperformed GPT-4o by achieving higher element recognition rate values in three out of five applications (Bludit, KanBoard and ExpressCart). In contrast, GPT-4o performed slightly better than DeepSeek on MediaWiki and PrestaShop. In particular, the element recognition rate achieved for ExpressCart was significantly higher than for the other applications. In this particular application, DeepSeek achieved 94\% element recognition Rate while GPT-4o reached 90\%. They performed both significantly better than the average element recognition Rate which was reported to be 73.4\% across all applications. 
GPT-4o and DeepSeek showed the lowest performance in PrestaShop, with element recognition rate values of 64\% and 61.2\% respectively. 
On the other hand, GPT-4o and DeepSeek performed best in ExpressCart application, reaching 90\% and 94\% element recognition rate.

It is important to emphasize that all POs produced by both LLMs exhibit proper class naming conventions, syntactically correct Java code with all necessary imports, and correctly defined constructors.

\begin{tcolorbox}[boxrule=0pt,frame hidden,sharp corners,enhanced,borderline north={1pt}{0pt}{black},borderline south={1pt}{0pt}{black},boxsep=2pt,left=2pt,right=2pt,top=2.5pt,bottom=2pt]
\textbf{RQ\textsubscript{3} (Issue categories):} The main difficulties arise in identifying method return types, with most errors due to missing return types (often defaulting to void) or methods that should return the same PO to enable chaining; the two LLMs behave similarly in this respect.
\end{tcolorbox}

\subsection{Threats to Validity}

Although the experiments yielded promising results, it is important to review potential threats to validity in order to correctly interpret the findings. These threats are discussed in the section below.

\subsubsection{Internal Validity}
Since the evaluation involved manual comparison of POs, we relied on a dataset of manually implemented ones that had been previously developed by other researchers and made publicly available. 
We consider these implementations reliable and used them the ground truth to assess the LLM's accuracy. 
However, this still introduces a potential validity threat, as the reference implementations may contain omissions or inaccuracies that could affect the reliability of our assessment and, consequently, the validity of our findings. 
Moreover, the manual validation process introduces a subjective component. 
The manual classification of LLM-generated outputs introduces subjectivity, as evaluators may interpret naming conventions differently. 
To mitigate this, multiple authors reviewed and discussed ambiguous cases until consensus was reached. 
All evaluators have substantial prior experience with PO design and implementation, which reduced the risk of systematic bias. 
Since the readability of POs is not an objective of this study, disagreements related to naming conventions do not affect our findings and are therefore not considered problematic.


Since the ground-truth POs represent one of multiple possible valid designs, structurally different yet functionally correct generations may be classified as deviations. We mitigated this by reviewing extra elements manually; several were found to implement valid, complementary interactions. Hence, our reported accuracy figures should be interpreted as a conservative lower bound.

Another potential threat to validity is data leakage from the pre-training of LLMs. 
Since LLMs are trained on extensive public data, some code structures or PO patterns used in our study may already exist in their training corpus. 
In such cases, models might reproduce patterns they have seen before rather than generate novel solutions, potentially leading to an overestimation of their actual ability to automate PO generation in unseen applications. 
As the exact training data are unknown, this remains a potential threat to our study's validity.

\subsubsection{External Validity} 


The experiments were conducted using only two LLMs due to local processing limitations constraints and the inability of smaller open-source models to complete the task. 
Consequently, our study focused on GPT-4o and DeepSeek, accessed via their APIs. 
While both are state-of-the-art models, their performance may not represent all existing LLMs, whose results can vary based on training data and architecture. 
Therefore, other models might outperform them.


The performance of LLMs in automating PO creation strongly depends on the provided prompt. We observed that both keyword choice and input structure significantly affected the output. Although we explored several configurations to optimize results, many other prompting strategies remain untested. These unexplored alternatives could yield better performance, making the limited set of prompts evaluated a potential threat to the validity of our work.



\subsubsection{Conclusion Validity}

Conclusion validity may be affected by the small number of LLMs and applications tested. The observed differences between GPT-4o and DeepSeek might be influenced by model-specific biases or the characteristics of the selected benchmarks rather than general trends. 
As a result, while the findings provide valuable insights into LLM performance on PO generation, they should be interpreted as indicative rather than conclusive.
When we began the study, we expected LLMs to struggle particularly with determining the correct return type. 
Unlike simple element classification, this requires understanding an element's role within the application flow, especially for navigational components that trigger page transitions. 
With only a single HTML file as input, the model must infer the correct target PO without access to the broader context, such as routing logic or related POs. 
A central question in our evaluation is whether LLMs can generalize these relationships from isolated HTML snapshots. 
In large applications, it is not feasible to include the full codebase in a single prompt due to context limitations, making it essential to assess how well LLMs perform under partial access to the web app. 
Understanding these limitations is key to judging their scalability and practical use, but is left for future work.
\section{Discussion}\label{sec:discussion}

The results obtained from our empirical study on the effectiveness of LLMs in automating page object (PO) generation indicate that ChatGPT-4o and DeepSeek perform comparably well. 
Both models achieved similar levels of accuracy and element recognition rates in the generated POs. Our study establishes an initial benchmark for automated PO generation with LLMs. While accuracy remains moderate, the consistency of syntactic correctness and high element recognition demonstrate tangible progress toward replacing template-based approaches with semantic generation.
The following section discusses these findings, along with their implications.

\head{Comparison of General-Purpose and Code-Oriented LLMs}
By employing GPT-4o and DeepSeek in five distinct applications, it can be concluded that LLMs are capable of generating effective POs. 
In most cases, the generated POs successfully captured the majority of GUI elements necessary for testing the application's functionality, with Element Recognition Rate values frequently exceeding 70\%.

In majority of the cases DeepSeek showed better performance in both accuracy and element recognition comparisons. 
However, usually there was not a large difference between two LLMs. 
Across all applications, DeepSeek demonstrated a tendency to produce more extensive POs than GPT-4o, incorporating a greater number of elements and methods.
Another characteristic observed in DeepSeek was to generate some methods only with the signature but without implementation, which also seemed different than GPT-4o.

It is important to point out that, in this research we employed DeepSeek Coder~\cite{guo2024deepseekcoder}, which is a variant of the DeepSeek model. 
This model has been fine-tuned and optimized for software development and code generation tasks. 
Thus, using it likely contributed to its stronger performance in both accuracy and Element Recognition Rate. 
Consequently, the choice of this particular model may have strengthened DeepSeek's tendency to produce more comprehensive POs compared to the more general-purpose LLM GPT-4o.

Although DeepSeek tended to generate a larger number of extra elements compared to GPT-4o, it is not yet clear whether these additions are truly beneficial for testing. 
In this study, we manually examined the extra methods by reviewing each PO individually. 
Many of them appeared to be reasonable and aligned with the basic functionality of the pages. 
However, to properly assess their practical value, more comprehensive test suites would be required, built upon POs that extend beyond core functionality to capture a broader range of behaviors and interactions. 
The contribution of these extra elements to functional coverage and fault detection remains an open question and should be investigated in future work. 

\head{Open Challenges in Structural Reasoning and Navigation Inference}
Both models exhibited similar challenges, particularly in correctly inferring the return types of navigation methods. 
This suggests that relying solely on static HTML and task instructions within a zero-shot prompting setup limits the models' ability to infer inter-object relationships. 
Without access to dynamic behaviors, such as user flows or page transitions, LLMs struggle to distinguish actions that lead to new pages from those that remain within the same view. 
As a result, incorrect assumptions about method return types frequently caused inconsistencies between the generated POs and the ground truth.

This limitation is especially evident in the naming of return types for navigational methods. 
In some cases LLMs were not able to identify that the method should navigate to another page, and instead they generated those methods with \texttt{void} return types. 
In some other cases, they recognized that a navigational action exists and attempt to redirect to another PO class. 
To be able to accurately define these methods, they must infer the target PO class name solely based on cues from the given HTML file,  such as link text or attribute values. 
If the provided HTML contains vague or non-descriptive link names, the model can be misled, resulting in incorrect or overly generic class names that do not correspond to the actual ones.
Even in cases where LLMs managed to produce contextually close and reasonable naming, there were cases in which the LLM failed to exactly match the expected class name. 
This most likely occurs due to subtle naming conventions or nuances that LLMs could not infer without awareness of the broader web app context. 

On the other hand, in cases where the LLMs successfully generated entirely correct navigational methods, the links within the HTML files were typically named almost identically to the target PO class names. 
This suggests that such naming consistency served as the primary cue the models relied upon to infer the correct PO naming. 
Consequently, the structure and semantics of the HTML input files play a crucial role in determining model performance.
Despite these shared shortcomings, each LLM exhibited distinct coding behaviors and stylistic variations in PO generation. 
These differences are likely attributable to disparities in their underlying training data, architectural biases, and prompt interpretation strategies. 
Overall, our findings highlight the limitations of zero-shot prompting; more sophisticated prompting techniques, such as chain-of-thought reasoning or few-shot examples, may significantly influence outcomes and warrant further investigation in future work.

\head{Impact of Input Size and Context Limitations}
The comparison between prompted HTML files and the completeness of the generated POs shows that lengthy HTML pages, even if not structurally complex, often challenge LLMs in capturing the full contextual relationships needed for complete page representations. 
For instance, in pages containing sidebars or menus with many similar links, the models frequently generated only a subset of elements, omitting others and thus lowering overall PO accuracy. 

This behavior can be attributed to several factors. 
First, when the combined size of the prompt and output nears the model's context window limit, generation may be truncated before all elements are produced—particularly for long pages with repetitive structures. 
Second, transformer-based models naturally avoid excessive repetition, often generating a few representative instances of recurring patterns rather than reproducing every occurrence explicitly.

Additionally, the phrasing of the prompt and the imposed generation constraints likely influenced these outcomes. 
If the instructions did not explicitly require exhaustive element coverage or the token budget was insufficient, the model may have prioritized conveying the overall structure over full completeness. 
This behavior highlights the importance of our preprocessing step, where HTML files were cleaned by removing headers and scripts, enabling the model to focus on relevant content and reducing premature truncation.

Across the five applications from different domains, we found no consistent correlation between application type and the accuracy or element recognition rate achieved by the LLMs. 
This indicates that structural complexity, repetition, and prompt design had greater impact on performance than domain characteristics. 
A larger and more domain-balanced dataset would be needed to confirm this observation. 
Future work should also explore iterative or conversational refinement, allowing the LLM to incrementally adjust generated POs based on developer feedback.

\section{Related work}\label{sec:related-work}

Several works in the literature, including recent ones based on LLMs, have explored various aspects of web testing automation~\cite{10.1145/3735553, sasazawa2025webpageclassificationusing,2020-Biagiola-ICST,2019-Biagiola-FSE-Diversity,2019-Biagiola-FSE-Dependencies,2026-Kanaththage-ICST}. 
However, when it comes to POs, most studies primarily focus on assessing their effectiveness or maintainability~\cite{leotta2020family,leotta_pageobject_study}, while only a very limited number of approaches have explicitly addressed the problem of automatically generating them~\cite{2016-Stocco-SQJ,7372274}.

The empirical study conducted by Leotta et al.~\cite{LeottaCRT13} indicates that test suites developed using a programmable approach (i.e., Selenium WebDriver) require higher initial development effort but results in lower maintenance effort compared to those created with a capture-and-replay approach (i.e., Selenium IDE). 
This outcome can be largely attributed to the adoption of the PO design pattern, which effectively decouples test logic from application logic. 

Subsequently, Leotta et al.~\cite{leotta_pageobject_study} further investigated the impact of applying the PO pattern on the maintainability of automated test suites. 
Their empirical results show that encapsulating GUI interaction logic within PO classes reduces overall maintenance effort by decoupling test scripts from structural changes in the application. 
This study reinforces the practical value of the PO pattern and underlines the importance of adopting structured design principles in test automation.

More recently, the same authors conducted controlled experiments~\cite{leotta2020family} to compare test suites developed with and without the PO pattern. 
Their analysis considered both the initial design effort and the long-term benefits associated with POs. 
The results show that, although introducing POs entails a significant upfront cost, a clear break-even point emerges as the test suite scales: when the suite grows to approximately ten times its original size, the use of POs becomes substantially more advantageous in terms of maintainability and robustness. 
Overall, these studies highlight that while POs require greater initial effort, they provide considerable benefits in scalability and maintainability for large and evolving test suites.

To the best of our knowledge, only two studies have addressed the automatic generation of POs. 
The first, Apogen~\cite{2016-Stocco-SQJ}, automates PO creation by combining dynamic crawling, clustering, and static analysis. 
It reverse-engineers the application under test, groups similar pages, builds a state-based model, and converts it into Java Page Objects following the Selenium Page Factory pattern. 
Unlike approaches that generate only class skeletons, Apogen produces complete POs with WebElement instances and methods for navigation and form submission, but does not employ LLMs.
The second work~\cite{7372274} introduces a prototype for automatically generating POs and PO-based test suites through feedback-directed random testing. 
It decouples test code from web pages by generating POs and was evaluated on seven real-world applications, achieving code coverage between 23.3\% and 90.8\% while revealing numerous HTML and runtime errors. 
Despite their promise, existing approaches~\cite{2016-Stocco-SQJ,7372274} remain research prototypes that often require manual correction of generated abstractions. 
In contrast, this work is the first to explore the use of LLMs for PO generation, aiming to improve automation and quality while reducing the need for human intervention.

\section{Conclusions and Future Work}\label{sec:conclusions}

This paper investigates the potential of LLMs to automate Page Object (PO) creation, an essential yet time-consuming task in web testing. 
By comparing LLM-generated POs with manually developed ones across multiple applications, we assessed their accuracy and element recognition rate against a curated ground truth. 
Our findings show that models such as GPT-4o and DeepSeek Coder can substantially reduce the manual effort required for PO development, though challenges persist in handling return types, navigation methods, and extraneous element generation. 
Notably, some of these additional methods may enhance GUI coverage, indicating that LLMs can capture meaningful patterns overlooked in manual design and potentially support test suite augmentation.

Future work will explore enhanced prompting strategies that explicitly encode page relationships, evaluate additional open-source models, extend this baseline study with few-shot and chain-of-thought prompting to assess contextual improvements, and using visual information~\cite{2020-Bajammal-TSE,2018-Stocco-FSE} and multi-modal LLMs. In addition, we plan to involve practitioners in user studies to provide more realistic insights into code quality~\cite{2021-Ricca-NEXTA,2021-Ricca-SOFSEM,2023-Ricca-QUATIC}, the usability of LLM-generated artifacts~\cite{2025-Ricca-IST}, the required effort for adjustments~\cite{leotta2020family}, and their actual impact on coverage and productivity. 


\bibliographystyle{ieeetr}
\bibliography{paper}

\end{document}